\documentclass[12pt]{article}
\usepackage{amsmath,amssymb}
\usepackage{latexsym,graphicx,epsfig,cite}

\def\beq{\begin{equation}}
\def\eeq{\end{equation}}
\def\bea{\begin{eqnarray}}
\def\eea{\end{eqnarray}}

\def\l#1{{\lambda}_{#1}}

\def \mij {m_{ij}}
\def \mjk {m_{jk}}
\def \mik {m_{ik}}
\def \mm {m_{\tilde G}}
\def \mi {m_{\nu_i}}
\def \mj {m_{\ell_j}}
\def \mk {m_{\ell_k}}

\def \mmn {m_{\tilde \nu_{i}}}
\def \mml {m_{\tilde \ell_{j}}}

\def\l {\lambda }

\def\lsim{\mathrel{\rlap{\raise 2.5pt \hbox{$<$}}\lower 2.5pt\hbox{$\sim$}}}
\def\gsim{\mathrel{\rlap{\raise 2.5pt \hbox{$>$}}\lower 2.5pt\hbox{$\sim$}}}

\newcommand{\sfermion}{{\tilde f}}
\newcommand{\gravitino}{{\tilde G}}

\renewcommand{\Re}{{\rm Re\thinspace}}

\allowdisplaybreaks %!!!!
\begin{document}

\pagestyle{empty}
\begin{flushright}
NORDITA-2008-51\\
CAVENDISH-HEP-2008-12\\
DAMTP-2008-88
\end{flushright}
\vspace*{5mm}
\begin{center}
{\large {\bf Gravitino Dark Matter and the Flavour Structure of
R-violating Operators}} \\
\vspace*{1cm}
{\bf N.-E. Bomark$^{1}$, S.\ Lola$^2$, P.\ Osland$^{1,3}$}
and {\bf A.R.\ Raklev$^{4,5}$} \\
\vspace{0.3cm}
$^1$ Department of Physics and Technology, University of Bergen, N-5020 Bergen,
Norway \\
$^2$ Department of Physics, University of Patras, GR-26500 Patras, Greece \\
$^3$ NORDITA, SE-10691 Stockholm, Sweden \\
$^4$ Department of Applied Mathematics and Theoretical Physics, University of Cambridge,
Cambridge CB3 0WA, UK \\
$^5$ Cavendish Laboratory, University of Cambridge, Cambridge CB3 0HE, UK

\vspace*{2cm}
{\bf ABSTRACT}
\end{center}
\vspace*{2mm}

We study gravitino dark matter and slow gravitino decays within the
framework of R-violating supersymmetry, with particular emphasis on
the flavour dependence of the branching ratios and the allowed
R-violating couplings. The dominant decay modes and final state
products turn out to be very sensitive to the R-violating
hierarchies. Mixing effects can be crucial in correctly deriving the
relative magnitude of the various contributions, particularly for
heavy flavours with phase space suppression. The study of the strength
of different decay rates for the gravitino is also correlated to
collider signatures expected from decays of the Next-to-Lightest
Supersymmetric Particle (NLSP) and to single superparticle production.

%\vspace*{2cm}

%\vfill\eject

\setcounter{page}{1}
\pagestyle{plain}

\pagebreak

\section{Introduction}

Recently, there has been renewed interest in the possibility of having
gravitino dark matter within the framework of R-violating
supersymmetry \cite{BM,LOR}, which occurs if the gravitino decays are
slow enough for its lifetime to be larger than the age of the universe
\cite{TY,CM}. This is an exciting possibility that
allows supersymmetric dark matter, even if the symmetries of the
fundamental theory result in an unstable Lightest Supersymmetric
Particle (LSP)~\cite{Rpar,Hall:1983id,barb}.

This is what happens if, in addition to the
couplings that generate the fermion and Higgs masses
\begin{equation}
\mu H_1 H_2 + m^e L_{i}\bar{E}_{j} H_1
+ m^d Q_{i}\bar{D}_{j} H_1
+ m^u Q_{i}\bar{E}_{j} H_2,
\label{mssm}
\end{equation}
we also have R-violating couplings of the form
\begin{equation}
h L_i H_2 + \lambda L_{i}L_{j}{\bar{E}}_{k}
+\lambda ^{\prime }L_{i}Q_{j}{\bar{D}_{k}}
+\lambda ^{\prime \prime }{\bar{U}_{i}}{\bar{D}_{j}}{\bar{D}_{k}}.
\label{Rviol}
\end{equation}
In the above, $H_{1,2}$ are the Higgs superfields, $L(Q)$ are the
left-handed lepton (quark) doublet superfields, and ${\bar{E}}$
(${\bar{D}},{\bar{U}}$) are the corresponding left-handed singlet
fields. The first three couplings in (\ref{Rviol}) violate lepton
number, while the fourth violates baryon number.

The stricter bounds on R-violating operators come from proton
stability, and R-parity \cite{fayet}, which forbids all lepton and
baryon number violating operators, is one of the possible
solutions. However, this is not the only symmetry that can guarantee
proton stability; baryon or lepton parities
\cite{IR,LR} can also exclude the simultaneous presence of dangerous
$LQ\bar{D}$ and $\bar{U}\bar{D}\bar{D}$ couplings~\cite{SMVIS}.
Experimental constraints from the non-observation of modifications to
Standard Model rates, or of possible exotic
processes~\cite{constraints} also impose additional
bounds\footnote{Additional strong constraints can be obtained from the
observation of NLSP decays to a gravitino LSP, with a photon or lepton
plus missing energy signature~\cite{CPW}.}.

R-violating supersymmetry results in a very rich phenomenology. In the
presence of the additional operators, the NLSP can decay into
conventional Standard Model particles. The missing energy signature of
the Minimal Supersymmetric Standard Model (MSSM)
\cite{HabKan} is substituted by multi-lepton and/or multi-jet events.
In addition to the consequences for collider searches, R-violation
implies that any gravitinos that have been thermally produced after a
period of inflation, are also unstable.

Gravitinos have three main decay modes: via tree-level three-body
decays to fermions~\cite{CM}, via two-body decays to neutrino and
photon due to neutrino--neutralino mixing~\cite{TY,BM}, and via
one-loop decays to neutrino and photon, generated by the trilinear
couplings \cite{LOR} \footnote{For heavy gravitinos, there is also the
possibility of producing massive gauge bosons. However, for trilinear
couplings and the range of parameters considered here, these
contributions are subdominant.}. In all three cases, the very large
suppression $1/M_p$ of the gravitino vertex, where $M_p$ is the
reduced Planck scale, plus additional suppression from phase space,
mixing and loop factors, respectively, result in large gravitino
lifetimes. For a wide set of R-parity violating couplings and
gravitino masses these exceed the age of the universe. Moreover, the
photon flux from these decays could be able to explain the apparent
excess in the extragalactic diffuse gamma-ray flux in the re-analysis
of the EGRET data~\cite{EGRET,reEGRET}.

The branching ratios for gravitino decays are
sensitive to the flavour structure of the R-violating operators.
In the case of {\cal O} (GeV) gravitinos, the presence of
tau or bottom quarks in the final state significantly enhances the
branching ratio of radiative decays with respect to the tree-level
ones, while for ``super-light'' gravitinos, as in \cite{Zw},
gravitinos are essentially stable with respect to the three-body
decays. Moreover, in the case of non-zero $\lambda ^{\prime \prime }
\bar{U}_{3} \bar{D}_j \bar{D}_k$ only --- with a top quark final state ---
gravitinos lighter than $m_t$ have a maximal stability, modulo mixing
effects, which we will discuss in a subsequent section\footnote{For an
operator of the form $\lambda ^{\prime} L_i Q_3 \bar{D}_k$ this
argument does not hold, since, when we pass from superfields to
component fields the $L_iQ_3$ part can become $\ell_i t$ or $\nu_i
b$.}.

In \cite{LOR}, gravitino decays were studied for $LL\bar{E}$ operators
that give rise to both loop and tree-level decays, with a tau or a
muon in the loop. Here, we extend the results in the following way:
\begin{itemize}
\item[(i)] We look at flavour effects in more detail,
making the link with fermion mass hierarchies. Within this framework
we comment on the relative magnitudes for bilinear and trilinear
R-violation and what are the implications for gravitino decays.
\item[(ii)] We extend the discussion to all 45 $LL\bar{E}$,
$LQ\bar{D}$ and $\bar{U}\bar{D}\bar{D}$ operators, paying particular
attention to the different features of the various decay modes and
possible bounds from gamma-ray measurements.
% (for instance, for $LQ\bar{D}$ operators, two-body gravitino decays to a
%lepton and either a pion or a kaon may occur; this does not happen
%with $\bar{U}\bar{D}\bar{D}$ operators, since the final states are all
%either particles or anti-particles).
\item[(iii)] We consider possible implications of mixing effects,
which in certain cases can be quite significant. For instance, for the
$\bar{U}_3 \bar{D}_j \bar{D}_k$ operator, the expected decay depends
very sensitively on the right quark mixing (for which little
information is available).
\item[(iv)] We link the above with probes of R-parity violation at the LHC,
in particularly NLSP decays, which may yield interesting signatures.
\end{itemize}

We begin in Section~\ref{sec:rad} by describing the various modes of
gravitino decays with trilinear couplings and the calculation of the
resulting extragalactic diffuse photon flux. In
Section~\ref{sec:flavstruc} we discuss possible flavour structures for
R-parity violating operators, before we look at the consequences for
gravitino decays in Section~\ref{sec:flavdcy}, with particular
attention to bounds from gamma-ray measurements. We continue with the
corresponding prospects for hadron colliders in
Section~\ref{sec:colliders}, before concluding in
Section~\ref{sec:conclusion}.

%%%%%%%%%%%%%%%%%%%%%%%%%%%%%%%%%%%%%%%%%%%%%%%%%%%%%%%%%%%%%%%%%%%%%%%%
\section{Gravitino Decays}
\label{sec:rad}
\setcounter{equation}{0}
%%%%%%%%%%%%%%%%%%%%%%%%%%%%%%%%%%%%%%%%%%%%%%%%%%%%%%%%%%%%%%%%%%%%%%%%

As already discussed, trilinear R-violating operators may cause
gravitinos to decay via two different channels:
\begin{itemize}
\item
Via two-body radiative loop decays to neutrino and photon
(Fig.~\ref{Fig:feyn})~\cite{LOR}.
\item
Via tree-level decays to fermions (Fig.~\ref{Fig:3-body})~\cite{CM}.
\end{itemize}

\begin{figure}[h]
\refstepcounter{figure}
\label{Fig:feyn}
\addtocounter{figure}{-1}
\begin{center}
\setlength{\unitlength}{1cm}
\begin{picture}(14.0,7.0)
\put(0.,0.)
{\mbox{\epsfysize=7cm
\epsffile{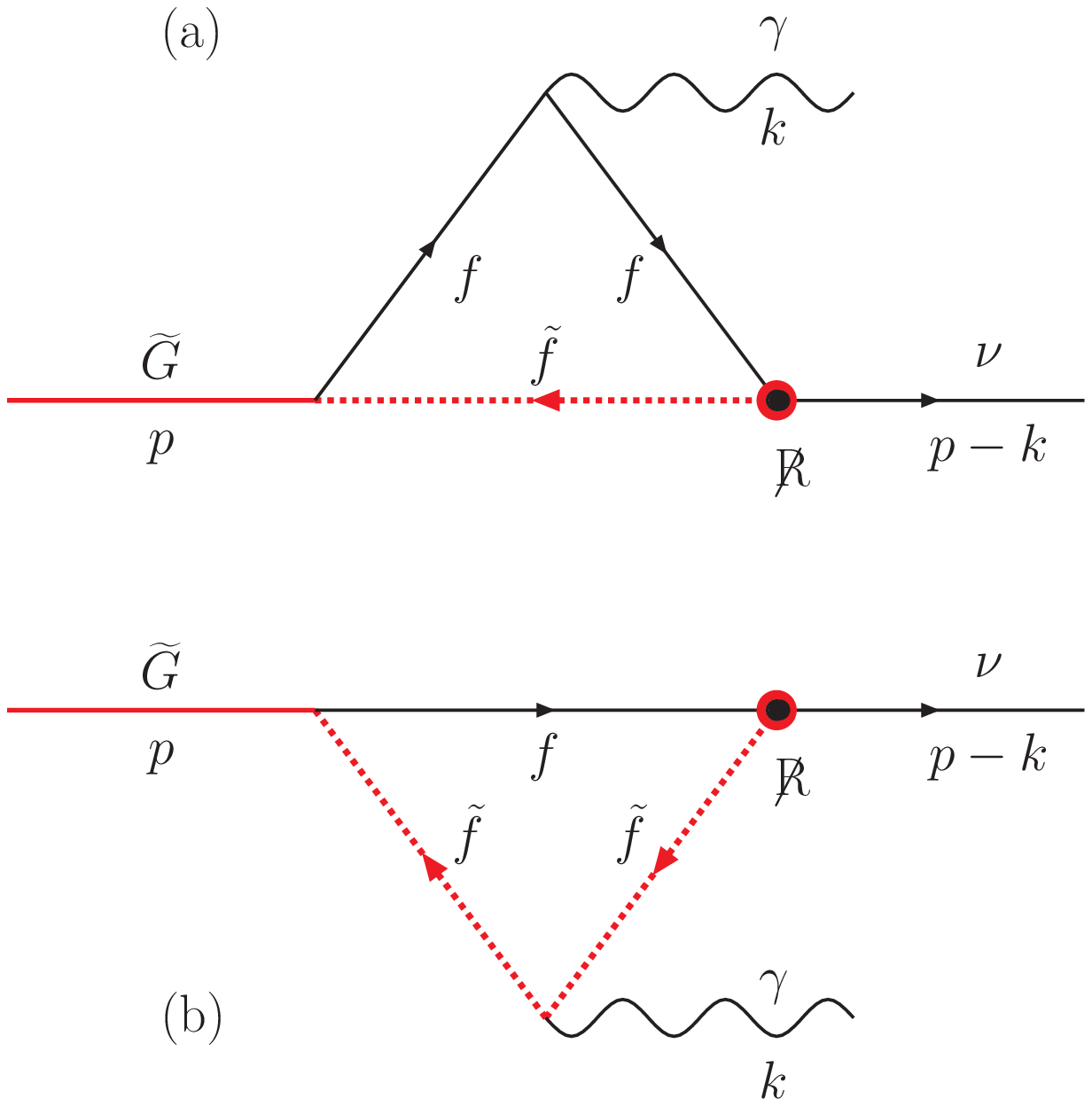}}}
\put(7.5,1.7)
{\mbox{\epsfysize=3.0cm
\epsffile{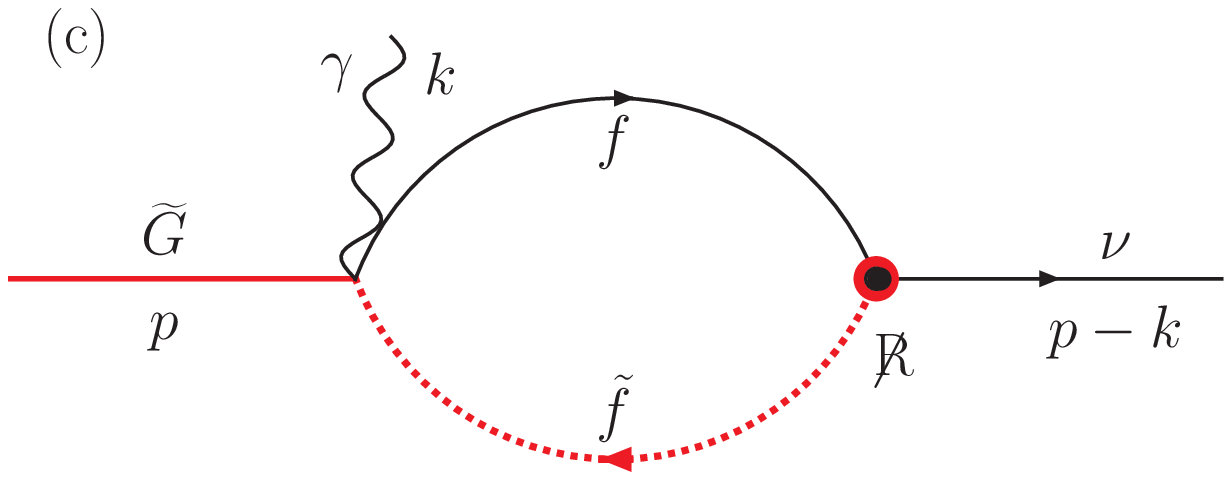}}}
\end{picture}
%\vspace*{-4mm}
\caption{\it Basic set of Feynman diagrams for radiative gravitino decay,
shown for (s)fermion loops. In the case of (s)quarks, the neutrino is
coupled to down-type quark to preserve SU(2) invariance. Arrows
denote flow of fermion number for left-chiral fields.}
\end{center}
\end{figure}

\begin{figure}[h]
\refstepcounter{figure}
\label{Fig:3-body}
\addtocounter{figure}{-1}
\begin{center}
\setlength{\unitlength}{1cm}
\begin{picture}(14.0,4.5)
\put(3.5,0.)
{\mbox{\epsfysize=5cm
\epsffile{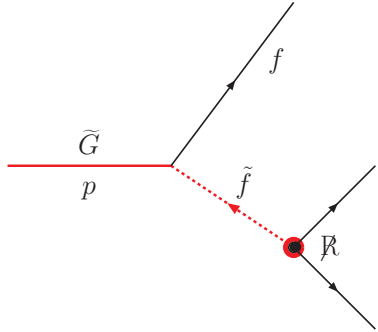}}}
\end{picture}
%\vspace*{-4mm}
\caption{\it Three-body decay of a gravitino via an R-parity violating coupling.
There are three contributing diagrams where the sfermion carries any
one of the three indices $i$, $j$ and $k$ of the corresponding
operators.}
\end{center}
\end{figure}

The decay rates have been presented in detail in the original
references, and for completeness are briefly summarised in Appendices
A and B, respectively. For light gravitino masses and appropriate
fermions in the loop the radiative decays may dominate. Indeed, as we
shall see, even when the three-body decay involving an intermediate
sfermion $\tilde{f}$ is well above the kinematical threshold at
$2m_f$, the radiative dominance is still present. The behaviour of the
decay rates is controlled by the mass dependence of the decay width:
for the three-body decay $\Gamma_{\tilde G}\propto m_{\tilde G}^7$,
while for the radiative decay $\Gamma_{\tilde G}\propto m_{\tilde G}$
at low gravitino masses. The latter occurs since the gravitational
coupling compensates for the relatively high loop mass by its
increasing strength for higher loop momenta. Because of the helicity
structure of the couplings, the two-body decay width is also strongly
dependent on the mass of the fermion in the loop, $\propto m_l^2$ at
low gravitino masses, implying significantly shorter lifetimes for
dominant third generation couplings.

To constitute a realistic dark matter candidate, the gravitino
lifetime should exceed the age of the universe. Moreover, the photon
flux from gravitino decays has to be consistent with observations. The
diffuse extra-galactic gamma ray flux of energy $E$ from the gravitino
decays is described by a integral over red-shift $z$ given by
\cite{Ibgam}
\begin{equation}
F(E)=E^2\frac{dJ}{dE}
=\frac{2E^2}{\mm}C_\gamma\int^\infty_1 dy \frac{dN_\gamma}{d(Ey)}
\frac{y^{-3/2}}{\sqrt{1+\kappa y^{-3}}},
\label{eq:photon_flux}
\end{equation}
where $y= 1+z$ and  $dN_\gamma/dE$ is the gamma
ray spectrum from the gravitino decay. Here
\begin{equation}
C_\gamma=\frac{\Omega_{\tilde G}\rho_c}
{8\pi\tau_{\tilde G} H_0 \Omega_M^{1/2}}
\quad {\rm and} \quad \kappa=\frac{\Omega_\Lambda}{\Omega_M}.
\end{equation}
For the radiative gravitino decay $dN_\gamma/dE = \delta(E-\mm/2)$ and
Eq.~(\ref{eq:photon_flux}) simplifies to \cite{BM}
\begin{equation}
F(E)=E^2\frac{dJ}{dE}={\rm BR}(\tilde G\to\gamma\nu)
C_\gamma(1+\kappa x^3)^{-1/2}x^{5/2}\theta(1-x),
\label{eq:photon_flux_2body}
\end{equation}
where $x=2E/m_{\tilde G}$. In the case of three-body decays the
hadronization of the produced particles and the resulting photon
spectrum have been calculated using PYTHIA 6.4
\cite{Sjostrand:2006za}. The photons from the three-body decays come mostly
from internal bremsstrahlung off leptons and from $\pi^0$ decays.

Using the original EGRET analysis \cite{EGRET}, with a power law
description of the extragalactic flux as
\begin{equation}
E^2\frac{dJ}{dE}=1.37\cdot 10^{-6}\left(\frac{1~{\rm GeV}}{E}\right)^{0.1}
{\rm GeV~cm^{-2}~sr^{-1}~s^{-1}},
\label{eq:egret}
\end{equation}
in the energy range 30~MeV to 100~GeV, severe bounds on gravitino
decays and thus on the allowed combinations of gravitino masses and
R-violating couplings can be derived. For comparison, predictions for
photonic spectra from gravitino decays through neutrino--neutralino
mixing, and also possible antimatter signatures of gravitino dark
matter, have recently been studied in~\cite{Ibgam}
and~\cite{Ishiwata:2008cu}.

%%%%%%%%%%%%%%%%%%%%%%%%%%%%%%%%%%%%%%%%%%%%%%%%%%%%%%%%%%%%%%%%%%%%%%%%
\section{Flavour Structure and Hierarchies of R-violating Operators}
\label{sec:flavstruc}
\setcounter{equation}{0}
%%%%%%%%%%%%%%%%%%%%%%%%%%%%%%%%%%%%%%%%%%%%%%%%%%%%%%%%%%%%%%%%%%%%%%%%

The implication of radiative gravitino decays as compared to the
tree-level ones, clearly depends on the flavour structure of the
R-violating operators involved. For higher generations the radiative
decay widths become larger and the tree-level diagrams suppressed due
to limited phase space. Most phenomenological studies assume a single
operator-dominance. This can be motivated by the fact that the Yukawa
couplings that generate fermion masses also have large hierarchies.
However, in principle, one may try to relate R-violating hierarchies
to those of fermion masses \cite{MODELS,ELR}, using models with family
symmetries. When exact, the latter allow only the third generation
fermions to become massive, while the remaining masses are generated
by the spontaneous breaking of this symmetry (see below). If $R$
parity is violated, couplings with different family charges will also
appear with different powers of the family symmetry-breaking
parameter, and thus with different magnitudes.

Moreover, one would have to appropriately take into account mixing
effects. Indeed, even with the common assumption of single
$R$-violating operator dominance, this would be true only for the
basis of current eigenstates for quarks and leptons, while, in the
mass-eigenstate basis, there would be several operators corresponding
to the original dominant one in the current basis. In addition, the
fact that there are strict bounds on some operators, implies that
mixing effects may in given models generate additional bounds on
couplings that at a first glance look less constrained. This has been
analysed in detail in \cite{ELR}, where it was shown that in theories
with strong correlations between operators (such as left-right
symmetric models), the effects can be particularly significant.

The starting point in such considerations, is to assume a $U(1)$
flavour symmetry, with the charges of the Standard Model fields
denoted as in Table~\ref{table:charges}.

\begin{table}[h]
\begin{center}
\begin{tabular}{|c|cccccccc|}
\hline
& $Q_{i}$ & $\bar{U}_i$
& $\bar{D}_i$ &
$L_i$ &
$\bar{E}_i$
& $\bar{N}$ & $H_{2}$ & $H_{1}$ \\
\hline
U(1) & $\alpha _{i}$ & $\beta _{i}$ & $\gamma _{i}$ & $c_{i}$ &
$d_{i}$ & $e_{i}$ & $-\alpha _{3}-\beta _{3}$ & $w$
\\ \hline
\end{tabular}
\caption{\label{table:charges}
\it Notation for possible U(1) charges of the various Standard Model
fields, where $i$ is a generation index.}
\end{center}
\end{table}

The flavour charge of $H_2$ is chosen so that the
operator that generates the top quark mass ($Q_3 \bar{U}_3 H_2$) has a
zero U(1) flavour charge and thus is allowed at zeroth order, as it
should be, since the top quark is significantly heavier than the
rest. The remaining matrix elements may be generated when the U(1)
symmetry is spontaneously broken \cite{FN,IR2} by fields $\theta, \;
\bar{\theta}$ that are singlets of the Standard Model gauge group,
with U(1) charges that are in most cases taken to be $\pm 1$,
respectively. For instance, for $\alpha_{i} =
\beta_{i}$ and $|\alpha_{3} -\alpha_{2}| = \pm 1$ as in \cite{IR}, the charm
mass comes about by a term
%$Q_2 \bar{U}_2 H_2 \left (\frac{\langle\theta\rangle}{M}\right)$
$Q_2 \bar{U}_2 H_2 \left (\langle\theta\rangle/M\right)$
or
%$Q_2 \bar{U}_2 H_2 \left(\frac{\langle\bar{\theta}\rangle}{M}\right)$
$Q_2 \bar{U}_2 H_2 \left(\langle\bar{\theta}\rangle/M\right)$
where $M$ is the heavy scale of the theory.

One may generalise the above to non-abelian flavour symmetries, and,
as an example, the following mass matrices have been proposed
\cite{KR}\ :
\[
M^\text{up} \propto \left(
\begin{array}{ccc}
0 & \epsilon ^{3} & \epsilon ^{3} \\
\epsilon ^{3} & \epsilon ^{2} & \epsilon^2 \\
\epsilon ^{3} & \epsilon^2 & 1
\end{array}
\right), ~M^\text{down} \propto \left(
\begin{array}{ccc}
0 & \bar{\epsilon} ^{3} & \bar{\epsilon} ^{4} \\
\bar{\epsilon} ^{3} & \bar{\epsilon} ^{2} & \bar{\epsilon}^2 \\
\bar{\epsilon} ^{3} & \bar{\epsilon}^2 & 1
\end{array}
\right), ~M^{\ell} \propto \left(
\begin{array}{ccc}
0 & \bar{\epsilon} ^{3} & \bar{\epsilon} ^{4} \\
\bar{\epsilon} ^{3} & \bar{\epsilon} ^{2} & \bar{\epsilon}^2 \\
\bar{\epsilon} ^{3} & \bar{\epsilon}^2 & 1
\end{array}
\right).
\label{mm}
\]
When diagonalising these matrices, the fermion mass hierarchies and
mixing are well reproduced for appropriate values of $\epsilon$,
$\epsilon \sim \bar{\epsilon}^2 \sim 0.04$. In general, in the models
appearing in the literature, the relative flavour charges in
Table~\ref{table:charges} and thus the exact structure of the mass
matrices are determined by the GUT multiplet structure (and the
requirement that particles in the same GUT multiplet have the same
charge). Nevertheless, in all cases, the observed fermion hierarchies
require smaller charges for the operators of the higher generations
(zero for the top Yukawa mass terms, but also for the bottom and tau
in a supersymmetric model with large $\tan\beta$). This implies that,
independently of the specific flavour and GUT structure of the theory,
{\em and unless extra fields with a non-zero flavour charge are
involved in the generation of R-violating couplings} \cite{ELR},
operators that contain fields of the third generation should be
naturally larger.

One has also to worry about the overall suppression of the R-violating
couplings with respect to the dominant Yukawa ones. However, this may
arise either from a small $\tan\beta$ in supersymmetric models, from
the form of the K\"ahler potential, or from additional, model
dependent, features of the theory that may involve extra fields and
symmetries.

Along these lines, one may also understand how it could be possible to
only have dominant $\bar{U}_3 \bar{D}_j\bar{D}_k$ operators. The
obvious step, to also ensure the absence of any unacceptable proton
decay, is to first eliminate lepton-number violating operators by
imposing a lepton triality, under which the fields transform as
\beq
Z_3: (Q,\bar{U},\bar{D},L,\bar{E},H_1,H_2)
\rightarrow (1,1,1,a,a^2,1,1).
\eeq
This allows only the baryon-number-violating operators and the mass
terms, while forbidding lepton-number-violating ones\footnote{A
flavour-dependent generalisation of this symmetry has been discussed
in \cite{LR}. In that scenario, consistent solutions were found
containing only a subclass of operators violating lepton number
($LL{\bar E}$) and baryon number (${\bar U}{\bar D} {\bar D}$). Thus
it is possible to have both lepton and baryon number violation without
disturbing proton stability.}. In this construction bilinear
R-violation would also be disallowed.

To allow only lepton-number violating operators, we could work instead
with a baryon triality, such as in \cite{IR}
\beq
Z_3: (Q,\bar{U},\bar{D},L,\bar{E},H_1,H_2)
\rightarrow (1,a^2,a,a^2,a^2,a^2,a).
\eeq
Such a baryon triality would allow for bilinear R-violation. However,
one may also envisage different structures where the symmetries forbid
an $L H_2$ term while allowing trilinear lepton-number violating
operators. An example of this is given by
\beq
Z_3: (Q,\bar{U},\bar{D},L,\bar{E},H_1,H_2)
\rightarrow (1,a,1,1,1,1,a^2).
\eeq
It is interesting to observe that in this case the term $\mu H_1 H_2$
would also be forbidden. This is due to charge correlations that arise
from the above requirements, plus the need to allow Yukawa couplings
that generate fermion masses. In this case, the $\mu$-term would have
to arise either radiatively \cite{muRad}, or through the K\"ahler
potential \cite{GuiMas}. The $\mu$ term could also be generated
within the framework of the NMSSM \cite{NMSSM}, via a singlet field
with appropriate charge; in which case a term $S L H_2$ would also be
allowed. Baryon number violating operators would be allowed at
subdominant orders, due to a term $S S \bar{U}\bar{D}\bar{D}$ which is
significantly suppressed; moreover, this is not the complete picture,
since to explain fermion mass hierarchies one would have to introduce
flavour dependent charges, which could further suppress R-violating
operators, particularly for the lighter generations that are dangerous
for proton decay (see discussion below).

From the above, it is clear that whether bilinear or trilinear R-violation
dominates is directly linked to the symmetries of the underlying theory,
and phenomenological information would be a valuable probe of this
symmetry structure.

Would these considerations be sufficient to understand the structure of the
R-violating operators on the basis of positive experimental results? As
already discussed, even in the case of one dominant operator, for fermions in
the basis of current eigenstates, mixing effects will induce non-zero
coefficients for related operators in the basis of mass eigenstates. These
will be suppressed by the mixing parameters with respect to the dominant
operator, but will not be zero, and this may affect phenomenological and
cosmological predictions. We should also keep in mind that experiments only
provide information on the Cabibbo-Kobayashi-Maskawa (CKM) quark mixing matrix
$V^\text{CKM}=V_{u}^{L\dagger }V_{d}^{L}$ \cite{Cabibbo:yz}, and that one can
construct theoretical models where the left quark mixing is in either the up
or the down sector, or both. Similarly, lepton mixing comes from the product of
matrices of charged leptons and neutrinos, with the additional complication
that, for the latter, we have the possibility of both Dirac and Majorana mass
terms (the recent neutrino data indicate the existence of neutrino masses and
contain the possibility that right-handed neutrinos do exist). For instance,
in the above mass matrices from~\cite{KR}, the quark mixing is given by
\[
V_{u}^{L,R}\approx \left(
\begin{array}{ccc}
1 & \epsilon & \epsilon ^{3} \\
-\epsilon & 1 & \epsilon^2 \\
\epsilon ^{3} & -\epsilon^2 & 1
\end{array}
\right) ,\;\;V_{d}^{L,R}\approx \left(
\begin{array}{ccc}
1 & \bar{\epsilon} & \bar{\epsilon} ^{4} \\
-\bar{\epsilon} & 1 & \bar{\epsilon}^2 \\
\bar{\epsilon} ^{4} & -\bar{\epsilon}^2 & 1
\end{array}
\right).
\label{eq:mixmat}
\]

Due to this mixing, an R-violating operator is in fact a sum of
terms. For instance, the mixing matrices above would lead to the
following interesting mixings:
\begin{equation}
\begin{array}{l}
(\bar{U}_{3}\bar{D}_i\bar{D}_j)^{\prime } =
\bar{U}_{3}\bar{D}_i\bar{D}_j
- \epsilon^2
\bar{U}_{2}\bar{D}_i\bar{D}_j + \epsilon^3
\bar{U}_{1}\bar{D}_i\bar{D}_j +\ldots\\
(L_{1}Q_{3}\bar{D}_{3})^{\prime } = L_{1}Q_{3}\bar{D}_{3} -\bar{\epsilon}^2
L_{1}Q_{3}\bar{D}_{2}+ \bar{\epsilon}^{4} L_{1}Q_{3}\bar{D}_{1}+\ldots
\end{array}
\label{opmix}
\end{equation}
These mixings are particularly important since the dominant couplings
here have massive final states. As we shall see, mixing also opens up
for final states forbidden by the gauge symmetry of the
couplings. However, more generically, we observe the following:
\begin{itemize}
\item[(i)] The right-handed quark mixing (relevant for $\bar{U}$
and $\bar{D}$) is essentially not constrained by the data. Therefore,
in a model with left-right asymmetric mass matrices, one could also
imagine a theory with a minimal mixing in the right-handed sector, in
which case a dominant $\bar{U}_{3}\bar{D}_i\bar{D}_j$ flavour would be
the only relevant one, and a gravitino with $m_\gravitino<M_t$ would
be essentially stable.
\item[(ii)] For the left quark mixing (relevant for $Q$), we know the
numerical values from $V_{CKM}$ (where for instance the 2-3 mixing is
a factor of $\approx 0.04$). Thus, a coupling $\lambda^{\prime} L_3
Q_3\bar{D}_3$, would in principle also imply the coupling $0.04
\lambda^{\prime} L_3 Q_2 \bar{D}_3$.
\item[(iii)] The left lepton mixing (relevant for $L$) is constrained
by the lepton data (large 1-2 and 2-3 mixing, and small 1-3 mixing).
\end{itemize}
We see that there are several flavour choices that can lead to
significant effects on the decays under discussion, particularly in
the cases where the available phase space is limited. This will be
explored in the next Section.

%%%%%%%%%%%%%%%%%%%%%%%%%%%%%%%%%%%%%%%%%%%%%%%%%%%%%%%%%%%%%%%%%%%%%%%%
\section{Flavour Effects in Gravitino Decays}
\label{sec:flavdcy}
\setcounter{equation}{0}

\subsection{Flavour Effects for $LL\bar{E}$ Operators}

From the nine R-violating $LL\bar{E}$ operators, six can potentially
give rise to both loop and tree-level decays (a common flavour in
$\bar{E}$ and one of the $L$ fields is needed to form the loop):
\bea
L_{2,3} L_1 \bar{E}_1, \quad L_{1,3} L_2 \bar{E}_2, \quad L_{1,2}L_3\bar{E}_3,
\label{bothdec}
\eea
while three have only three-body decays
\bea
L_{2} L_3 \bar{E}_1, \quad L_{1} L_3 \bar{E}_2, \quad L_{1} L_2 \bar{E}_3.
\label{only-tree}
\eea
The cases with a muon or a tau in the loop were discussed in
\cite{LOR}. For an electron in the loop, the photonic gravitino decays
are very suppressed due to the electron mass, and the tree-level
decays dominate unless the gravitino becomes extremely light. This is
demonstrated in Figure~\ref{Fig:max1}, where we plot $\lambda_{\max}$,
the maximum allowed coupling, versus the gravitino mass, assuming a
common slepton mass of 200~GeV. In doing so, we demand that:
\begin{itemize}
\item[(i)] there is one dominant coupling,
\item[(ii)] the gravitinos can be dark matter, with a lifetime of at
least $10$ times the current age of the universe and that
\item[(iii)] photon production from the gravitino decays, as calculated by
Eq.~(\ref{eq:photon_flux}), is consistent with the bounds on the photon
spectrum given in Eq.~(\ref{eq:egret}).
\end{itemize}

%%%%%%%%%%%%%%%%%%%%%%%%%%%%%%%%%%%%%%%%%%%%%%%%%%%%%%%%%%%%%%%%%%%%%%%%
\begin{figure}[h]
\refstepcounter{figure}
\label{Fig:max1}
\addtocounter{figure}{-1}
\begin{center}
\setlength{\unitlength}{1cm}
\begin{picture}(15.0,7.5)
\put(1.5,0.)
{\mbox{\epsfysize=7.5cm
\epsffile{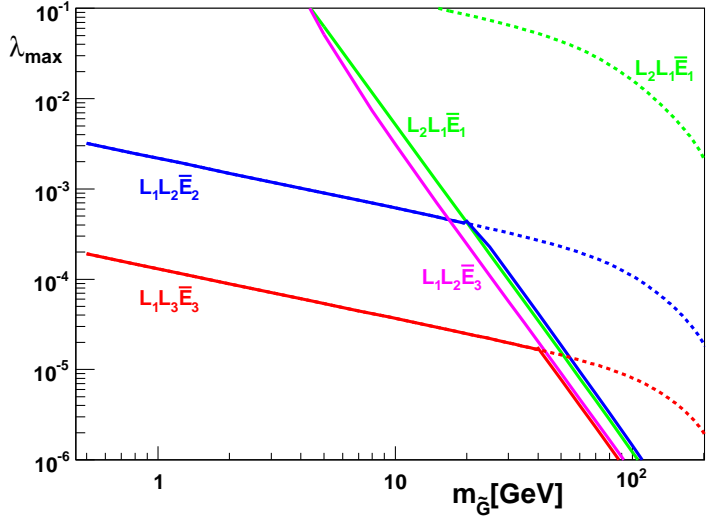}}}
\end{picture}
%\vspace*{-4mm}
\caption{\it Maximum value $\lambda_{\max}$ of R-violating couplings versus
gravitino mass, for $LL\bar{E}$ operators. Bounds shown as dashed
lines are when considering radiative loop decays only, solid lines
include photons from three-body decays. The sparticle masses are 200~GeV.}
\end{center}
\end{figure}
%%%%%%%%%%%%%%%%%%%%%%%%%%%%%%%%%%%%%%%%%%%%%%%%%%%%%%%%%%%%%%%%%%%%%%

We see that while the photon flux from two-body loop decays puts
strong bounds on couplings that lead to loops with muons (blue) or
taus (red), the couplings with electron loops (green) are only
affected by the three-body decay photons down to very small gravitino
masses. For the couplings (\ref{only-tree}) with no loop diagrams the
bounds are thus correspondingly weak, and follow the bound for
$L_{2}L_1\bar{E}_1$. As expected, the neutrino flavour has no effect
on the bounds from the radiative decay, so results for e.g.\
$L_{1}L_2\bar{E}_2$ and $L_{3}L_2\bar{E}_2$ are virtually identical,
save for minute differences near the slepton threshold.

It is also interesting to note that in the terms with only three-body
decays in (\ref{only-tree}), there is always the possibility for
tau production in the final state. Indeed, for $L_{1}L_2\bar{E}_3$
an SU(2) singlet $\tau$ is always produced if kinematically allowed,
while for $ L_{2} L_3\bar{E}_1$ and $L_{1} L_3 \bar{E}_2$ an SU(2)
doublet $\tau$ is produced, unless the gravitino mass becomes lower or
comparable to the tau. In this case the factor $L_{1,2}L_3$ would
only contribute to the tree-level decay via the $\nu_{\tau} e$ or
$\nu_{\tau} \mu$ term. This is observed in Figure~\ref{Fig:max1},
where in the three-body dominated region, bounds on e.g.\ $L_1 L_3
\bar{E}_3$ are stricter than the bounds on $L_1 L_2\bar{E}_2$ and
$L_2 L_1 \bar{E}_1$, due to the extra photons from the tau decay. One
can also notice that the bound on $L_2 L_1 \bar{E}_1$ is slightly
better than on $L_1 L_2\bar{E}_2$; this is due to more bremsstrahlung
from electrons than from muons in the final state.

%%%%%%%%%%%%%%%%%%%%%%%%%%%%%%%%%%%%%%%%%%%%%%%%%%%%%%%%%%%
\subsection{Flavour effects in $L Q \bar{D}$ operators}
%%%%%%%%%%%%%%%%%%%%%%%%%%%%%%%%%%%%%%%%%%%%%%%%%%%%%%%%%%

Out of the 27 R-violating $LQ\bar{D}$ operators, only the following
nine can potentially give rise to both loop and tree-level decays (a
common flavour in $Q$ and $\bar{D}$ is needed to form the loop):
\bea
L_{1,2,3} Q_1 \bar{D}_1, \quad L_{1,2,3} Q_2\bar{D}_2,
\quad L_{1,2,3} Q_3 \bar{D}_3,
\label{bothdec-2}
\eea
while the remaining 18 have only three-body decays.

In Figure~\ref{Fig:max2} we show a comparison of the partial lifetime
for the loop and tree-level decays for the second and third
generation. We choose $L_3$, but this has little
significance. Comparing to the results for the $LL\bar{E}$ operators
in~\cite{LOR}, we observe that with a $b$ quark instead of a $\tau$ in
the loop, radiative decays dominate over the three-body ones for a
significantly wider range of gravitino masses, up to 40 GeV, for the
same sparticle masses (200~GeV). This arises both due to the higher
fermion mass in the loop, but also due to the two bottom masses in the
final state. The coupling $L_iQ_3\bar{D}_3$ gives rise to either
$\ell_i t \bar{b}$ or $\nu_i b \bar{b}$, and the first term is
forbidden by phase space up to high gravitino masses, which can be
seen as a bump in the $L_3Q_3\bar{D}_3$ three-body lifetime near
threshold.\footnote{Close to the $b\bar{b}$ threshold at $\sim 10$~GeV
hadronization effects will become important for the three-body
decay. This is not considered here as the two-body decay clearly
dominates in this mass range.}

%%%%%%%%%%%%%%%%%%%%%%%%%%%%%%%%%%%%%%%%%%%%%%%%%%%%%%%%%%%%%%%%%%%%%%%%
\begin{figure}[h]
\refstepcounter{figure}
\label{Fig:max2}
\addtocounter{figure}{-1}
\begin{center}
\setlength{\unitlength}{1cm}
\begin{picture}(15.0,7.5)
\put(1.5,0.)
{\mbox{\epsfysize=7.5cm
\epsffile{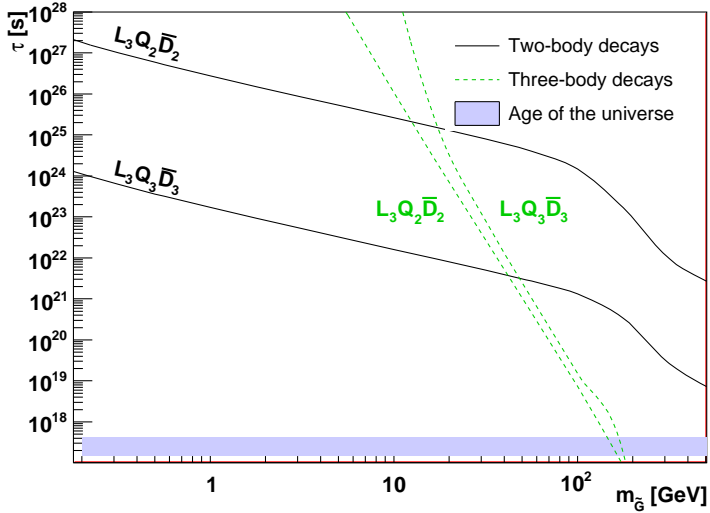}}}
\end{picture}
%\vspace*{-4mm}
\caption{\it Comparison of partial lifetime versus gravitino mass for
two-body loop decays and three-body tree-level decays for the
$L_2Q_3\bar{D}_3$ and $L_3Q_3\bar{D}_3$ couplings. Couplings have all been
set to $\lambda^\prime=0.001$.}
\end{center}
\end{figure}
%%%%%%%%%%%%%%%%%%%%%%%%%%%%%%%%%%%%%%%%%%%%%%%%%%%%%%%%%%%%%%%%%%%%%%

As in the previous subsection we can put constraints on the couplings
of the $LQ\bar{D}$ operators from gamma rays and gravitino lifetime,
as a function of the gravitino mass. The resulting bounds for
operators with both loop and tree-level decays are shown in
Figure~\ref{Fig:max3}. Due to the increased dominance of two-body
decays compared to the pure lepton operators, we have even stronger
coupling bounds, in particular for the $L_iQ_3\bar{D}_3$ couplings,
and there is now also a significant constraint on the first generation
loops, i.e.\ $L_iQ_1\bar{D}_1$, for low gravitino masses.

If the mass of the gravitino is close to the lightest possible meson
for one particular operator, we may no longer neglect hadronisation
effects from the formation of single mesons, as opposed to the QCD jet
interpretation of the quarks in the three-body decay. In the simplest
case we would have a two-body final state with a lepton and a meson,
such as a pion or a kaon, or even heavier mesons if allowed by the
structure of the R-violating operator and the mass of the
gravitino. For instance, the operator $L_3 Q_1 \bar{D}_1$ will lead to
$ \tau \pi^+ (\tau u\bar{d})$ or $\nu_\tau \pi^0 (\nu_\tau d
\bar{d})$, and similar considerations hold for other flavour
combinations. 

However, since the decay into single mesons is only relevant for low
gravitino masses, this issue can be neglected for operators allowing
loop decays. This is because the constraint from the loop decay to
photon and neutrino is in all cases a lot more stringent than the
constraint arising from the decay into mesons at these gravitino
masses.

For operators not permitting loop decays the situation is different,
but in the cases with light mesons the resulting gamma ray constraints
are so weak that other constraints on the couplings are more important
\cite{constraints}. Thus the only cases where decays into single mesons
can have some effect are for operators which give heavy mesons, {\it
i.e.}\ B or D mesons. In these cases there can be small modifications
on the constraints in the range of gravitino masses close to the heavy
quark masses, but the nature of heavy quarks as kinematically
equivalent to their corresponding mesons should limit this effect.

%%%%%%%%%%%%%%%%%%%%%%%%%%%%%%%%%%%%%%%%%%%%%%%%%%%%%%%%%%%%%%%%%%%%%%%%
\begin{figure}[h]
\refstepcounter{figure}
\label{Fig:max3}
\addtocounter{figure}{-1}
\begin{center}
\setlength{\unitlength}{1cm}
\begin{picture}(15.0,7.5)
\put(1.5,0.)
{\mbox{\epsfysize=7.5cm
\epsffile{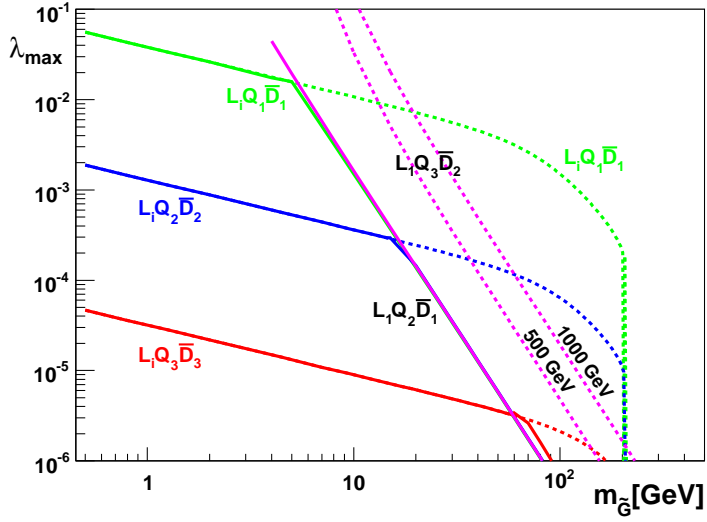}}}
\end{picture}
%\vspace*{-4mm}
\caption{\it Maximum value $\lambda_{\max}$ of R-violating couplings versus
gravitino mass, for $LQ\bar{D}$ operators.
The sparticle masses are 200~GeV, except where indicated.}
\end{center}
\end{figure}
%%%%%%%%%%%%%%%%%%%%%%%%%%%%%%%%%%%%%%%%%%%%%%%%%%%%%%%%%%%%%%%%%%%%%%

\subsection{Flavour effects in $\bar{U}\bar{D}\bar{D}$ operators}

In this case, we only have tree-level gravitino decays, and of particular
interest is the possibility of gravitino decays via a dominant
$\bar{U}_3\bar{D}_j\bar{D}_k$ operator. For light gravitinos, since top
production in the final state is kinematically forbidden, decays will arise
either
\begin{itemize}
\item[(i)] due to t-c mixing and other possible mixings,
\item[(ii)] or from four-body final states with an off-shell top quark and
possibly an off-shell $W$, and with at least one massive final state
particle (b-quark).
\end{itemize}
The first case is expected to dominate since the second is very
suppressed, and the dominant decay width should be a function of the
right-handed $\bar{U}_3-\bar{U}_2$ mixing. In this case
$\lambda_{\max}$ can be large, with interesting phenomenological
implications that we discuss in the next Section. Another interesting
feature of mixing is that it opens up gravitino decay channels that
were disallowed by the flavour structure of the superpotential, e.g.\
the possibility of two $b$ (or $\bar{b}$) in the final state.

Both of these effects are shown in Figure~\ref{Fig:udd}, where we plot
the partial lifetime for a selection of gravitino decay modes as a
function of gravitino mass. We assume a dominant coupling
$\lambda''_{312}=1.0$ that for low gravitino masses relies on mixing
effects in the decays. The coupling is chosen large to minimize the
lifetime. We illustrate a possible realization of mixing with the
mixing matrices in Eq.~(\ref{eq:mixmat}), taking $\epsilon=0.04$ and
$\bar\epsilon=0.20$. As expected, it is the t-c mixing that dominates
gravitino decays at low masses, and the gravitino is long lived enough
to be dark matter for a large range of masses. Only for gravitino
masses above 200~GeV, when the top production threshold has been
passed with good margin, do the top channels dominate and the
gravitino becomes disallowed as a dark matter candidate due to its
short lifetime.

%%%%%%%%%%%%%%%%%%%%%%%%%%%%%%%%%%%%%%%%%%%%%%%%%%%%%%%%%%%%%%%%%%%%%%%%
\begin{figure}[!h]
\refstepcounter{figure}
\label{Fig:udd}
\addtocounter{figure}{-1}
\begin{center}
\setlength{\unitlength}{1cm}
\begin{picture}(15.0,7.5)
\put(1.5,0.)
{\mbox{\epsfysize=7.5cm
\epsffile{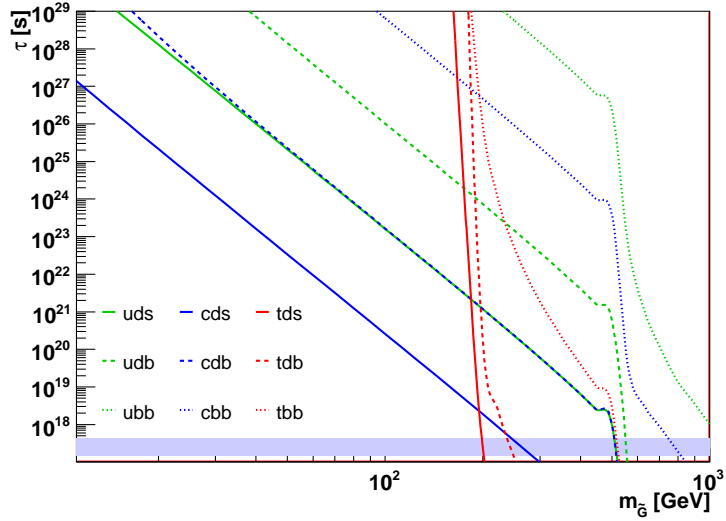}}}
\end{picture}
%\vspace*{-4mm}
\caption{\it Partial lifetime versus gravitino mass for gravitino decays
into various quark final states with $\lambda''_{312}=1.0$,
$\epsilon=0.04$ and $\bar\epsilon=0.20$. All squark masses have been
set to $500$~GeV.}
\end{center}
\end{figure}
%%%%%%%%%%%%%%%%%%%%%%%%%%%%%%%%%%%%%%%%%%%%%%%%%%%%%%%%%%%%%%%%%%%%%%

We find that changing between the three possible $\lambda''_{3jk}$
couplings only changes the relative importance of the down type quarks
in the gravitino decay, e.g.\ for $\lambda''_{313}=1.0$,
$\gravitino\to cdb$ is the dominant decay channel for low
masses. Among the channels that are closed in the absence of mixing,
we only show decays to $b$ quark pairs. The lighter pairs have very
similar behaviour to other light quark pairs. We see that the
probability of two $b$ quarks in the final state is negligible because
of the large suppression due to mixing and kinematics when compared to
other decay channels. Other choices for the mixing matrices only change
the relative importance of the different decay channels, not the
behaviour as a function of gravitino mass.

In Figure~\ref{Fig:maxUDD} we also show the resulting bounds on the
$\lambda^{\prime\prime}$ couplings when considering the photon spectrum as in
the previous subsections. We notice that the first two generations have a
log-linear behaviour in terms of the gravitino mass, with equal slopes.  The
difference in scaling is due to different squark masses.  With the same squark
mass, the two curves would be indistinguishable.  The importance of mixing
effects are again shown for the $\lambda''_{312}$ coupling: the opening up of
decays through mixing strengthens the bounds on that coupling.

%%%%%%%%%%%%%%%%%%%%%%%%%%%%%%%%%%%%%%%%%%%%%%%%%%%%%%%%%%%%%%%%%%%%%%%%
\begin{figure}[h]
\refstepcounter{figure}
\label{Fig:maxUDD}
\addtocounter{figure}{-1}
\begin{center}
\setlength{\unitlength}{1cm}
\begin{picture}(15.0,7.5)
\put(1.5,0.)
{\mbox{\epsfysize=7.5cm
\epsffile{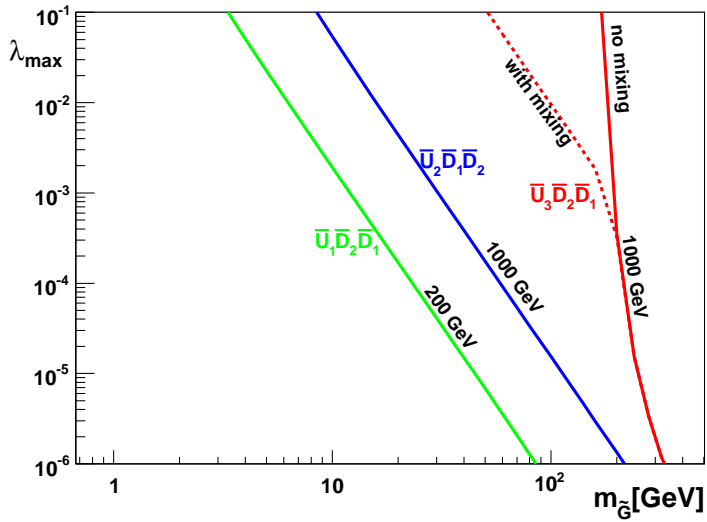}}}
\end{picture}
%\vspace*{-4mm}
\caption{\it Maximum value $\lambda_{\max}$ of R-violating couplings versus
gravitino mass, for $\bar{U}\bar{D}\bar{D}$ operators. The squark masses
are as indicated.}
\end{center}
\end{figure}
%%%%%%%%%%%%%%%%%%%%%%%%%%%%%%%%%%%%%%%%%%%%%%%%%%%%%%%%%%%%%%%%%%%%%%

In general, due to the structure of these operators we produce either
three quarks or three anti-quarks. If there is sufficient phase space,
one could imagine that we can end up with two-body final states with a
baryon and a meson for very light gravitino masses. However, we need
to keep in mind that the lightest flavours for
$\lambda^{\prime\prime}$, in particular $\lambda^{\prime\prime}_{112}$
and $\lambda^{\prime\prime}_{113}$, are extremely constrained from
double nucleon decay and neutron-antineutron oscillations
respectively~\cite{BarbMas}.

\section{Prospects for R-violation in colliders}
\setcounter{equation}{0}
\label{sec:colliders}

For R-violating couplings above $10^{-6}$ for 100 GeV sparticle
masses, and with a scaling that for most operators is a simple
proportionality relation, one would expect interesting signatures like
multi-lepton and/or multi-jet events in the final state of sparticle
production in a collider. Depending on the flavour of the R-violating
operator, the nature of the NLSP, and the respective $\lambda_{\max}$
that we found in the previous section, one would generically expect
either:
\begin{itemize}
\item[(i)] possible observable single superparticle productions, if $\lambda$
can be sufficiently large~\cite{Rpar,Hall:1983id,DR},
\item[(ii)] MSSM production of sparticle pairs followed by R-violating
decays of the NLSP, for the flavours where $\lambda_{\max}$ is smaller
than $\sim 10^{-2}$, or
\item[(iii)] no R-violatig decays of the NLSP inside detectors for very small
$\lambda_{\max}$ (smaller than $\sim 10^{-6}$), with some cross-over
region where displaced vertices could be observed.
\end{itemize}

From the results shown in the previous Section, the observation of single
sparticle production at the LHC is almost entirely excluded in the gravitino
dark matter scenario for operators that give loop decays with second or third
generation loop-particles, due to the strict bounds from gamma rays. For
dominant three-body decays the same conclusion holds unless the gravitino mass
is small ($m_{\tilde G}\lsim 10$~GeV).  Thus the possible astrophysical
observation of gravitino decays will have important consequences for LHC
expectations, and vice versa. It is worth noting that this conclusion, for the
case of dominant two-body decay, is only weakly dependent on the assumed
masses of the other sparticles, as can be seen from the insensitivity of
$\lambda_{\max}$ to large changes in intermediate sparticle mass, see Figure~5
of~\cite{LOR}.

For no operator do the constraints considered here eliminate the
possibility of seeing R-violating decays in colliders, but the
$L_iL_3\bar{E}_3$ and $L_iQ_3\bar{D}_3$ operators allow only a very
restricted coupling range for intermediate to high gravitino masses.
Indeed, even for couplings of the order of $10^{-6}$ it should be
possible to detect the R-violating NLSP decays~\cite{DR}.

The discovery of supersymmetry at the LHC and the reconstruction of a
neutralino NLSP has been shown to be possible~\cite{:1999fr} at the
same level or better than for R-parity conserving scenarios when one
considers the lepton number violating operators. This is due to the
numerous leptons expected in the final state. However, for the case of a
$\bar{U}\bar{D}\bar{D}$ operator, assumptions have to be made, either
for the production of additional leptons in the event from cascade
decays, or for heavy flavours that can be tagged. The heavier the
flavours, the better the detection prospects due to flavour tagging or
top reconstruction.

Decays of the NLSP are highly dependent on the combination of NLSP
flavour and dominant R-parity violating operator flavour. If these
flavours are the same, all NLSP decays should be rapid two-body decays
if kinematically allowed. In other scenarios, three or even four-body
decays are the leading decays, with resulting suppression due to phase
space and heavy virtual particles. For an
$\bar{U}_{3}\bar{D}_j\bar{D}_k$ operator, we have the following
particular implications:
\begin{itemize}
\item[(i)] we have the possibility of large R-violating coupling with
resulting resonant single stop production \cite{single-stop} or single
gluino top production \cite{gluino-top}. Moreover, for a large
$\lambda^{\prime \prime}_{3jk} \lambda^{\prime \prime}_{i3k}$ product,
one may observe interesting signatures in single top--bottom
production \cite{top-bottom}.
\item[(ii)] if the NLSP is a neutralino with a mass larger than the top,
it should have a rapid three-body decay with a top in the final state,
on the other hand, if the neutralino is lighter than the top, then it
should decay via either subdominant operators or mixing effects, which
may well enhance its decay rate enough for it to decay within the
detector, giving a displaced vertex.
\end{itemize}

Taken together this would imply the interesting possibility of
sparticle production via one operator, and decay via a different
one.

%%%%%%%%%%%%%%%%%%%%%%%%%%%%%%%%%%%%%%%%%%%%%%%%%%%%%%%%%%%%%%%%%%%%%%%%
\section{Conclusions}
\label{sec:conclusion}
\setcounter{equation}{0}
%%%%%%%%%%%%%%%%%%%%%%%%%%%%%%%%%%%%%%%%%%%%%%%%%%%%%%%%%%%%%%%%%%%%%%%%
We have studied slow gravitino decays originating from lepton or
baryon number violating operators in R-violating supersymmetry,
focusing on the flavour structure of the theory. We found that the
dominant decay modes, and thus the final state products are
particularly sensitive to the hierarchies of R-violating operators and
exhibit distinct correlations, which we have analysed. Already the
dominance of trilinear R-violating couplings over bilinear modes implies
the presence of symmetries that, among others, have interesting
implications for the $\mu$-term.

A more detailed study of the flavour dependence of the operators has
determined the ratio between (i) the tree-level gravitino decays to
three fermions and (ii) the two-body loop decays into a photon and a
neutrino, which in turn puts strong bounds on the maximal value of the
allowed R-violating couplings. Bounds from photon spectra are much
stricter than the ones from the requirement on the gravitino lifetime,
and thus strongly constrain the respective operators, particularly
$L_i L_3 \bar{E}_3$ and $L_i Q_3 \bar{D}_3$ that involve a $\tau$ and
a bottom-quark in the loop. On the other hand, for operators without
photonic decays larger coupling constants are possible, particularly
in the case of phase space suppressions due to the presence of heavy
fermions in the final state. Moreover, mixing effects turn out to be
crucial in correctly deriving the relative magnitude of the various
contributions, and play a significant role for decay modes with phase
space suppression and particularly for the ones generated by
$\bar{U}_3 \bar{D}_j\bar{D}_k$.

In all cases, the bounds on the R-violating couplings from the
cosmological requirements are compatible with visible signatures at
colliders, which can vary from single superparticle production (for
flavours where a larger coupling constant is allowed) to MSSM
production and R-violating decays (for the smaller
couplings). Particularly for the operator flavours that would lead to
predominantly photonic gravitino decays, giving strong constraints on
the couplings, interesting event properties such as vertex
displacement might be expected.

\appendix

\section{Photonic Gravitino Decays}
\setcounter{equation}{0}

The photonic decays of the gravitinos have been calculated
in~\cite{LOR}. The rate for the radiative decay $\tilde
G\to\gamma\nu$ with the loop fermion $f$ is given by
\begin{equation}
\Gamma=\frac{\alpha\lambda^2\,m_\gravitino}{2048\pi^4}\,
\frac{m_f^2}{M_p^2}
\overline{\left|{\cal F}\right|^2},
\end{equation}
where $M_p=(8\pi G_N)^{-1/2}=2.4\times10^{18}~\text{GeV}$ is the reduced
Planck mass, and\footnote{Here, we correct a minor error in that paper
due to a misprint in the gravitino spin-sum taken
from~\cite{Moroi:1995fs}, where the sign in Eq.~(4.31) should be
$(\not\! p + m_{3/2})$.}
\begin{align}
\overline{\left|{\cal F}\right|^2}
&=\frac{1}{12}|c_1|^2+\frac{2}{6}|c_2|^2+\frac{1}{6}\Re(c_1^*c_2),
\end{align}
with
\begin{align}
c_1&=2[(m_\gravitino^2-m_\sfermion^2+m_f^2)C_0^{(a)}+\Delta B_0^{(1)}],
\nonumber \\
c_2&=2[m_f^2 C_0^{(a)}+m_\sfermion^2\,C_0^{(b)}+\Delta B_0^{(2)}],
\label{eq:c1c2}
\end{align}
where, in the notation of {\tt LoopTools}
\cite{Hahn:1998yk,vanOldenborgh:1989wn}, we have
\begin{align} \label{Eq:finite_expressions}
C_0^{(a)}&=C_0(m_\gravitino^2,0,0,m_\sfermion^2,m_f^2,m_f^2), \nonumber \\
C_0^{(b)}&=C_0(m_\gravitino^2,0,0,m_\sfermion^2,m_f^2,m_\sfermion^2),
\nonumber \\
\Delta B_0^{(1)}&=2B_0(m_\gravitino^2,m_\sfermion^2,m_f^2)
-B_0(0,m_\sfermion^2,m_f^2)-B_0(0,m_f^2,m_f^2), \nonumber \\
\Delta B_0^{(2)}&=B_0(m_\gravitino^2,m_\sfermion^2,m_f^2)
-B_0(0,m_\sfermion^2,m_f^2).
\end{align}
The $C_0$ are three-point functions corresponding to
Fig.~\ref{Fig:feyn} (a) and (b), whereas the $\Delta B_0$ are finite
differences of two-point functions.

\section{Three-body Gravitino Decays}
\setcounter{equation}{0}

The three-body decays of gravitinos have been calculated in \cite{CM},
where extensive analytic formulas were derived. Here, we only comment
on the spin summed squared amplitudes, and refer to the original paper
for the full computation.

The full squared amplitude (summed over spins) for the gravitino decay
$\tilde G \rightarrow {\l_{ijk}}{\to} \nu_i \ell_j \bar \ell_k$ is the
sum of three individual squared amplitudes plus three interference
terms. These arise since the gravitino can couple to all the particles
involved in the R-violating operator. Then, for the case where the
gravitino couples to a neutrino and a sneutrino, one has
\begin{align}
\vert M_a \vert^2
&= \frac{1}{3}\frac{\l_{ijk}^2}{M_p^2 (m^2_{jk}-\mmn^2)^2}
(\mm^2-\mjk^2+\mi^2) (\mjk^2-\mj^2-\mk^2) \nonumber \\
&\times
\bigg ( \frac{(\mm^2+\mjk^2-\mi^2)^2}{4 \mm^2}-\mjk^2 \bigg ),
\label{amp1}
\end{align}
where $m_{jk}^2 = (p_j+p_k)^2$, with $p_{j,k}$ the four-momenta of the
respective particles. The remaining squared amplitudes $M_{b,c}$,
where the gravitino couples to the charged lepton of the doublet and
the singlet charged lepton respectively, are given by the same
formula, when substituting the appropriate flavours in the vertices
and the propagator. The interference terms are of the form
\begin{align}
2 \Re(M_a M^\dagger_{b})
&=\frac{1}{3} \frac{\l_{ijk}^2}
{ M_p^2 (m^2_{jk}-\mmn^2) (m^2_{ik}-\mml^2)} \bigg [
(\mik^2 \mjk^2 - \mm^2 \mk^2 - \mi^2 \mj^2) \nonumber \\
&\times
\bigg (
(\mm^2+\mk^2-\mi^2-\mj^2)
-\frac{1}{ 2 \mm^2}(\mm^2+\mjk^2-\mi^2)
(\mm^2+\mik^2-\mj^2) \bigg ) \nonumber \\
&+ \frac{1}{2} (\mij^2 - \mi^2 - \mj^2) (\mjk^2 - \mj^2 - \mk^2)
(\mik^2 - \mi^2 - \mk^2) \nonumber \\
&- \frac{\mi^2}{2} (\mjk^2 - \mj^2 - \mk^2)^2
- \frac{\mj^2}{2} (\mik^2 - \mi^2 - \mk^2)^2 \nonumber \\
& - \frac{\mk^2}{2} (\mij^2 - \mi^2 - \mj^2)^2 + 2 \mi^2 \mj^2 \mk^2 \bigg ].
\label{amp12}
\end{align}

For $LQ\bar{D}$ operators the results are similar, and found by
replacing the SU(2) doublet field $L$ by $Q$, and the SU(2) singlet
$\bar{E}$ by $\bar{D}$, and summing over colours. For
$\bar{U}\bar{D}\bar{D}$ operators we also have similar amplitudes and
interference terms. Again the contributions can be read off from
Eq.~(\ref{amp1}) and~(\ref{amp12}), modulo colour and symmetry factors
that arise from the possibility of two identical particles in the
final state.

%%%%%%%%%
{\bf Acknowledgements.} We thank the NORDITA program {\it ``TeV scale
physics and dark matter''}, for hospitality while part of this work
was carried out. We would like to thank C. Luhn and C. Savoy for very
useful comments. The research of SL is funded by the FP6 Marie Curie
Excellence Grant MEXT-CT-2004-014297. Participation in the European
Network MRTPN-CT-2006 035863-1 (UniverseNet) is also acknowledged.
The research of PO has been supported by the Research Council of
Norway. ARR acknowledges funding from the UK Science and Technology
Facilities Council (STFC).

%%%%%%%%%%%%%%%%%%%%%%%%%%%%%%%%%%%%%%%%%%%%%%%%%%%%%%%%%%%%%%%%%%%%%%


\begin{thebibliography}{99}
%%%%%%%%%%%%%%%%%%%%%%%%%%%%%%%%%%%%%%%%%%%%%%%%%%%%%%%%%%%%%%%%%%%%%%

\bibitem{BM}
%\cite{Buchmuller:2007ui}
%\bibitem{Buchmuller:2007ui}
W.~Buchmuller, L.~Covi, K.~Hamaguchi, A.~Ibarra and T.~Yanagida,
%``Gravitino dark matter in R-parity breaking vacua,''
JHEP {\bf 0703} (2007) 037
[arXiv:hep-ph/0702184].
%%CITATION = JHEPA,0703,037;%%

\bibitem{LOR}
%\bibitem{Lola:2007rw}
S.~Lola, P.~Osland and A.~R.~Raklev,
%``Radiative gravitino decays from R-parity violation,''
Phys.\ Lett.\ B {\bf 656} (2007) 83
[arXiv:0707.2510 [hep-ph]].
%%CITATION = PHLTA,B656,83;%%

\bibitem{TY}
%\cite{Takayama:2000uz}
%\bibitem{Takayama:2000uz}
F.~Takayama and M.~Yamaguchi,
%``Gravitino dark matter without R-parity,''
Phys.\ Lett.\ B {\bf 485} (2000) 388
[arXiv:hep-ph/0005214].
%%CITATION = PHLTA,B485,388;%%

\bibitem{CM}
G.~Moreau and M.~Chemtob,
%``R-parity violation and the cosmological gravitino problem,''
Phys.\ Rev.\ D {\bf 65}, 024033 (2002)
[arXiv:hep-ph/0107286].
%%CITATION = PHRVA,D65,024033;%%

\bibitem{Rpar}
For some of the earliest references
on the phenomenology of $R$-violating supersymmetry,
see: \\
%\cite{Zwirner:1984is}
%\bibitem{Zwirner:1984is}
F.~Zwirner,
%``Observable Delta B=2 Transitions Without Nucleon Decay In A Minimal
%Supersymmetric Extension Of The Standard Model,''
Phys.\ Lett.\ B {\bf 132} (1983) 103.
%%CITATION = PHLTA,B132,103;%%
%
%\cite{Ellis:1984gi}
%\bibitem{Ellis:1984gi}
J.~R.~Ellis, G.~Gelmini, C.~Jarlskog, G.~G.~Ross and J.~W.~F.~Valle,
%``Phenomenology Of Supersymmetry With Broken R-Parity,''
Phys.\ Lett.\ B {\bf 150} (1985) 142;
%%CITATION = PHLTA,B150,142;%%
%
%\cite{Ross:1984yg}
%\bibitem{Ross:1984yg}
G.~G.~Ross and J.~W.~F.~Valle,
%``Supersymmetric Models Without R-Parity,''
Phys.\ Lett.\ B {\bf 151} (1985) 375;
%%CITATION = PHLTA,B151,375;%%
%

%
%\cite{Dimopoulos:1988jw}
%\bibitem{Dimopoulos:1988jw}
S.~Dimopoulos and L.~J.~Hall,
%``Lepton and Baryon Number Violating Collider Signatures from
%Supersymmetry,''
Phys.\ Lett.\ B {\bf 207} (1988) 210.
%%CITATION = PHLTA,B207,210;%%

%\cite{Hall:1983id}
\bibitem{Hall:1983id}
L.~J.~Hall and M.~Suzuki,
%``Explicit R-Parity Breaking In Supersymmetric Models,''
Nucl.\ Phys.\ B {\bf 231} (1984) 419;
%%CITATION = NUPHA,B231,419;%%
%
%\cite{Dawson:1985vr}
%\bibitem{Dawson:1985vr}
S.~Dawson,
%``R-Parity Breaking In Supersymmetric Theories,''
Nucl.\ Phys.\ B {\bf 261} (1985) 297;
%%CITATION = NUPHA,B261,297;%%

\bibitem{barb}
For a review on R-violating supersymmetry, see:
R.~Barbier {\it et al.},
%``R-parity violating supersymmetry,''
Phys.\ Rept.\ {\bf 420}, 1 (2005)
[arXiv:hep-ph/0406039].
%%CITATION = PRPLC,420,1;%%

\bibitem{fayet}
%\cite{Fayet:1977yc}
%\bibitem{Fayet:1977yc}
P.~Fayet,
%``Spontaneously Broken Supersymmetric Theories Of Weak, Electromagnetic And
%Strong Interactions,''
Phys.\ Lett.\ B {\bf 69} (1977) 489.
%%CITATION = PHLTA,B69,489;%%

\bibitem{IR}
%\cite{Ibanez:1991hv}
%\bibitem{Ibanez:1991hv}
L.~E.~Ibanez and G.~G.~Ross,
%``Discrete gauge symmetry anomalies,''
Phys.\ Lett.\ B {\bf 260} (1991) 291;
%%CITATION = PHLTA,B260,291;%%
%\cite{Ibanez:1991pr}
%\bibitem{Ibanez:1991pr}
% L.~E.~Ibanez and G.~G.~Ross,
%``Discrete Gauge Symmetries And The Origin Of Baryon And Lepton Number
%Conservation In Supersymmetric Versions Of The Standard Model,''
Nucl.\ Phys.\ B {\bf 368} (1992) 3.
%%CITATION = NUPHA,B368,3;%%

\bibitem{LR}
%\cite{Lola:1993ip}
%\bibitem{Lola:1993ip}
S.~Lola and G.~G.~Ross,
%``Baryon and lepton number (non)conservation in supersymmetric theories,''
Phys.\ Lett.\ B {\bf 314} (1993) 336.
%%CITATION = PHLTA,B314,336;%%

\bibitem{SMVIS}
See
%\cite{Smirnov:1996bg}
%\bibitem{Smirnov:1996bg}
A.~Y.~Smirnov and F.~Vissani,
%``Upper bound on all products of R-parity violating couplings $\lambda'$ and
%$\lambda''$ from proton decay,''
Phys.\ Lett.\ B {\bf 380} (1996) 317
[arXiv:hep-ph/9601387],
%%CITATION = PHLTA,B380,317;%%
and references therein.
%\cite{Carena:1997wy}

\bibitem{constraints}
For a review of experimental constraints on
$R$-violating operators, see: \\
%\cite{Bhattacharyya:1997vv}
%\bibitem{Bhattacharyya:1997vv}
G.~Bhattacharyya,
%``A brief review of R-parity-violating couplings,''
arXiv:hep-ph/9709395;
%%CITATION = HEP-PH/9709395;%%
%
%\cite{Bhattacharyya:1996nj}
%\bibitem{Bhattacharyya:1996nj}
% G.~Bhattacharyya,
%``R-parity-violating supersymmetric Yukawa couplings: A mini-review,''
Nucl.\ Phys.\ Proc.\ Suppl.\ {\bf 52A} (1997) 83
[arXiv:hep-ph/9608415],
%%CITATION = NUPHZ,52A,83;%%
and references therein;
%
%\cite{Dreiner:1997uz}
%\bibitem{Dreiner:1997uz}
H.~K.~Dreiner,
%``An introduction to explicit R-parity violation,''
arXiv:hep-ph/9707435,
%%CITATION = HEP-PH/9707435;%%
published in {\it Perspectives on Supersymmetry},
ed. by G. Kane, World Scientific;
%
%\cite{Allanach:1999ic}
%\bibitem{Allanach:1999ic}
B.~C.~Allanach, A.~Dedes and H.~K.~Dreiner,
%``Bounds on R-parity violating couplings at the weak scale and at the GUT
%scale,''
Phys.\ Rev.\ D {\bf 60} (1999) 075014
[arXiv:hep-ph/9906209].
%%CITATION = PHRVA,D60,075014;%%

\bibitem{CPW}
M.~Carena, S.~Pokorski and C.~E.~M.~Wagner,
%``Photon signatures for low energy supersymmetry breaking and broken
%R-parity,''
Phys.\ Lett.\ B {\bf 430} (1998) 281
[arXiv:hep-ph/9801251].
%%CITATION = PHLTA,B430,281;%%

\bibitem{HabKan}
For an introduction to the phenomenology of the MSSM, see
%\cite{Nilles:1983ge}
%\bibitem{Nilles:1983ge}
H.~P.~Nilles,
%``Supersymmetry, Supergravity And Particle Physics,''
Phys.\ Rept.\ {\bf 110} (1984) 1;
%%CITATION = PRPLC,110,1;%%
%\cite{Haber:1984rc}
%\bibitem{Haber:1984rc}
H.~E.~Haber and G.~L.~Kane,
%``The Search For Supersymmetry: Probing Physics Beyond The Standard Model,''
Phys.\ Rept.\ {\bf 117} (1985) 75.
%%CITATION = PRPLC,117,75;%%

%\cite{EGRET}
\bibitem{EGRET}
P.~Sreekumar {\it et al.} [EGRET Collaboration],
%``EGRET observations of the extragalactic gamma ray emission,''
Astrophys.\ J.\ {\bf 494} (1998) 523
[arXiv:astro-ph/9709257].
%%CITATION = ASJOA,494,523;%%

%\cite{reEGRET}
\bibitem{reEGRET}
A.~W.~Strong, I.~V.~Moskalenko and O.~Reimer,
%``A new determination of the extragalactic diffuse gamma-ray background from
%EGRET data,''
Astrophys.\ J.\ {\bf 613} (2004) 956
[arXiv:astro-ph/0405441];
%%CITATION = ASJOA,613,956;%%
%\cite{Strong:2004de}
%\bibitem{Strong:2004de}
%A.~W.~Strong, I.~V.~Moskalenko and O.~Reimer,
%``Diffuse Galactic continuum gamma rays. A model compatible with EGRET data
%and cosmic-ray measurements,''
Astrophys.\ J.\ {\bf 613} (2004) 962
[arXiv:astro-ph/0406254].
%%CITATION = ASJOA,613,962;%%

\bibitem{Zw}
%\cite{Brignole:1998me}
%\bibitem{Brignole:1998me}
A.~Brignole, F.~Feruglio, M.~L.~Mangano and F.~Zwirner,
%``Signals of a superlight gravitino at hadron colliders when the other
%superparticles are heavy,''
Nucl.\ Phys.\ B {\bf 526} (1998) 136
[Erratum-ibid.\ B {\bf 582} (2000) 759]
[arXiv:hep-ph/9801329];
%%CITATION = NUPHA,B526,136;%%
%
%\cite{Brignole:1997sk}
%\bibitem{Brignole:1997sk}
A.~Brignole, F.~Feruglio and F.~Zwirner,
%``Signals of a superlight gravitino at e+ e- colliders when the other
%superparticles are heavy,''
Nucl.\ Phys.\ B {\bf 516} (1998) 13
[Erratum-ibid.\ B {\bf 555} (1999) 653]
[arXiv:hep-ph/9711516];
%%CITATION = NUPHA,B516,13;%%
%
%\cite{Brignole:1998uu}
%\bibitem{Brignole:1998uu}
% A.~Brignole, F.~Feruglio and F.~Zwirner,
%``Four-fermion interactions and sgoldstino masses in models with a
%superlight gravitino,''
Phys.\ Lett.\ B {\bf 438} (1998) 89
[arXiv:hep-ph/9805282].
%%CITATION = PHLTA,B438,89;%%

\bibitem{Ibgam}
A.~Ibarra and D.~Tran,
%``Gamma Ray Spectrum from Gravitino Dark Matter Decay,''
Phys.\ Rev.\ Lett.\ {\bf 100}, 061301 (2008)
[arXiv:0709.4593 [astro-ph]];
%%CITATION = PRLTA,100,061301;%%
%
%\cite{Ibarra:2008qg}
%\bibitem{Ibarra:2008qg}
A.~Ibarra and D.~Tran,
%``Antimatter Signatures of Gravitino Dark Matter Decay,''
JCAP {\bf 0807} (2008) 002
[arXiv:0804.4596 [astro-ph]].
%%CITATION = JCAPA,0807,002;%%

\bibitem{Sjostrand:2006za}
  T.~Sjostrand, S.~Mrenna and P.~Skands,
  %``PYTHIA 6.4 physics and manual,''
  JHEP {\bf 0605} (2006) 026
  [arXiv:hep-ph/0603175].
  %%CITATION = JHEPA,0605,026;%%

%\cite{Ishiwata:2008cu}
\bibitem{Ishiwata:2008cu}
K.~Ishiwata, S.~Matsumoto and T.~Moroi,
%``High Energy Cosmic Rays from the Decay of Gravitino Dark Matter,''
arXiv:0805.1133 [hep-ph];
%%CITATION = ARXIV:0805.1133;%%
%
%\cite{Ishiwata:2008cv}
%\bibitem{Ishiwata:2008cv}
K.~Ishiwata, S.~Matsumoto and T.~Moroi,
%``Cosmic-Ray Positron from Superparticle Dark Matter and the PAMELA
%Anomaly,''
arXiv:0811.0250 [hep-ph].
%%CITATION = ARXIV:0811.0250;%%

\bibitem{MODELS}
%\cite{BenHamo:1994bq}
%\bibitem{BenHamo:1994bq}
V.~Ben-Hamo and Y.~Nir,
%``Implications of horizontal symmetries on baryon number violation in
%supersymmetric models,''
Phys.\ Lett.\ B {\bf 339} (1994) 77
[arXiv:hep-ph/9408315];
%%CITATION = PHLTA,B339,77;%%
%
%\cite{Chamseddine:1995gb}
%\bibitem{Chamseddine:1995gb}
A.~H.~Chamseddine and H.~K.~Dreiner,
%``Anomaly Free Gauged R Symmetry In Local Supersymmetry,''
Nucl.\ Phys.\ B {\bf 458} (1996) 65
[arXiv:hep-ph/9504337];
%%CITATION = NUPHA,B458,65;%%
%
%\cite{Binetruy:1996xk}
%\bibitem{Binetruy:1996xk}
P.~Binetruy, S.~Lavignac and P.~Ramond,
%``Yukawa textures with an anomalous horizontal abelian symmetry,''
Nucl.\ Phys.\ B {\bf 477} (1996) 353
[arXiv:hep-ph/9601243];
%%CITATION = NUPHA,B477,353;%%
%
%\cite{Bhattacharyya:1998vw}
%\bibitem{Bhattacharyya:1998vw}
G.~Bhattacharyya,
%``Non-abelian flavour symmetry and R-parity,''
Phys.\ Rev.\ D {\bf 57} (1998) 3944
[arXiv:hep-ph/9707297];
%%CITATION = PHRVA,D57,3944;%%
%
%\cite{Binetruy:1997sm}
%\bibitem{Binetruy:1997sm}
P.~Binetruy, E.~Dudas, S.~Lavignac and C.~A.~Savoy,
%``Quark flavour conserving violations of the lepton number,''
Phys.\ Lett.\ B {\bf 422} (1998) 171
[arXiv:hep-ph/9711517].
%%CITATION = PHLTA,B422,171;%%

\bibitem{ELR}
%\cite{Ellis:1998rj}
%\bibitem{Ellis:1998rj}
J.~R.~Ellis, S.~Lola and G.~G.~Ross,
%``Hierarchies of R-violating interactions from family symmetries,''
Nucl.\ Phys.\ B {\bf 526} (1998) 115
[arXiv:hep-ph/9803308].
%%CITATION = NUPHA,B526,115;%%

\bibitem{FN}
C.~D.~Froggatt and H.~B.~Nielsen,
%``Hierarchy Of Quark Masses, Cabibbo Angles And CP Violation,''
Nucl.\ Phys.\ B {\bf 147} (1979) 277.
%%CITATION = NUPHA,B147,277;%%

\bibitem{IR2}
L.~E.~Ibanez and G.~G.~Ross,
%``Fermion masses and mixing angles from gauge symmetries,''
Phys.\ Lett.\ B {\bf 332} (1994) 100
[arXiv:hep-ph/9403338].
%%CITATION = PHLTA,B332,100;%%

\bibitem{KR}
S.~F.~King and G.~G.~Ross,
%``Fermion masses and mixing angles from SU(3) family symmetry,''
Phys.\ Lett.\ B {\bf 520}, 243 (2001)
[arXiv:hep-ph/0108112].
%%CITATION = PHLTA,B520,243;%%

\bibitem{muRad}
L.~J.~Hall, J.~D.~Lykken and S.~Weinberg,
%``Supergravity As The Messenger Of Supersymmetry Breaking,''
Phys.\ Rev.\ D {\bf 27} (1983) 2359.
%%CITATION = PHRVA,D27,2359;%%


\bibitem{GuiMas}
G.~F.~Giudice and A.~Masiero,
%``A Natural Solution to the mu Problem in Supergravity Theories,''
Phys.\ Lett.\ B {\bf 206}, 480 (1988).
%%CITATION = PHLTA,B206,480;%%


\bibitem{NMSSM}
%\cite{Fayet:1974pd}
%\bibitem{Fayet:1974pd}
P.~Fayet,
%``Supergauge Invariant Extension Of The Higgs Mechanism And A Model For The
%Electron And Its Neutrino,''
Nucl.\ Phys.\ B {\bf 90} (1975) 104;
%%CITATION = NUPHA,B90,104;%%
%
%\cite{Nilles:1982dy}
%\bibitem{Nilles:1982dy}
H.~P.~Nilles, M.~Srednicki and D.~Wyler,
%``Weak Interaction Breakdown Induced By Supergravity,''
Phys.\ Lett.\ B {\bf 120} (1983) 346;
%%CITATION = PHLTA,B120,346;%%
%
%\cite{Frere:1983ag}
%\bibitem{Frere:1983ag}
J.~M.~Frere, D.~R.~T.~Jones and S.~Raby,
%``Fermion Masses And Induction Of The Weak Scale By Supergravity,''
Nucl.\ Phys.\ B {\bf 222} (1983) 11;
%%CITATION = NUPHA,B222,11;%%
%
%\cite{Derendinger:1983bz}
%\bibitem{Derendinger:1983bz}
J.~P.~Derendinger and C.~A.~Savoy,
%``Quantum Effects And SU(2) X U(1) Breaking In Supergravity Gauge Theories,''
Nucl.\ Phys.\ B {\bf 237} (1984) 307;
%%CITATION = NUPHA,B237,307;%%
%
%\cite{Durand:1988rg}
%\bibitem{Durand:1988rg}
L.~Durand and J.~L.~Lopez,
%``Upper Bounds on Higgs and Top Quark Masses in the Flipped SU(5) x U(1)
%Superstring Model,''
Phys.\ Lett.\ B {\bf 217} (1989) 463;
%%CITATION = PHLTA,B217,463;%%
%
%\cite{Drees:1988fc}
%\bibitem{Drees:1988fc}
M.~Drees,
%``Supersymmetric Models with Extended Higgs Sector,''
Int.\ J.\ Mod.\ Phys.\ A {\bf 4} (1989) 3635;
%%CITATION = IMPAE,A4,3635;%%
%
%\cite{Ellis:1988er}
%\bibitem{Ellis:1988er}
J.~R.~Ellis, J.~F.~Gunion, H.~E.~Haber, L.~Roszkowski and F.~Zwirner,
%``Higgs Bosons in a Nonminimal Supersymmetric Model,''
Phys.\ Rev.\ D {\bf 39} (1989) 844.
%%CITATION = PHRVA,D39,844;%%

%\cite{Cabibbo:yz}
\bibitem{Cabibbo:yz}
N.~Cabibbo,
%``Unitary Symmetry And Leptonic Decays,''
Phys.\ Rev.\ Lett.\ {\bf 10} (1963) 531;\\
%%CITATION = PRLTA,10,531;%%
%\cite{Kobayashi:fv}
%\bibitem{Kobayashi:fv}
M.~Kobayashi and T.~Maskawa,
%``CP Violation In The Renormalizable Theory Of Weak Interaction,''
Prog.\ Theor.\ Phys.\ {\bf 49} (1973) 652.
%%CITATION = PTPKA,49,652;%%

%\cite{BarbMas}
\bibitem{BarbMas}
R.~Barbieri and A.~Masiero,
%``Supersymmetric Models With Low-Energy Baryon Number Violation,''
Nucl.\ Phys.\ B {\bf 267} (1986) 679.
%%CITATION = NUPHA,B267,679;%%

\bibitem{DR}
%\cite{Dreiner:1991pe}
%\bibitem{Dreiner:1991pe}
H.~K.~Dreiner and G.~G.~Ross,
%``R-parity violation at hadron colliders,''
Nucl.\ Phys.\ B {\bf 365} (1991) 597.
%%CITATION = NUPHA,B365,597;%%

%\cite{:1999fr}
\bibitem{:1999fr}
ATLAS collaboration,
``ATLAS detector and physics performance. Technical design report. Vol. 2''.
%%CITATION = ATLAS-TDR-15;%%

%\cite{Moroi:1995fs}
\bibitem{Moroi:1995fs}
T.~Moroi,
%``Effects of the gravitino on the inflationary universe,''
arXiv:hep-ph/9503210.
%%CITATION = HEP-PH/9503210;%%

\bibitem{single-stop}
%\cite{Berger:1999zt}
%\bibitem{Berger:1999zt}
E.~L.~Berger, B.~W.~Harris and Z.~Sullivan,
%``Single-top-squark production via R-parity-violating supersymmetric
%couplings in hadron collisions,''
Phys.\ Rev.\ Lett.\ {\bf 83}, 4472 (1999)
[arXiv:hep-ph/9903549].
%%CITATION = PRLTA,83,4472;%%

\bibitem{gluino-top}
%\cite{Chaichian:2000ux}
%\bibitem{Chaichian:2000ux}
M.~Chaichian, K.~Huitu and Z.~H.~Yu,
%``R parity violation in (t + anti-t) gluino production at LHC and
%Tevatron,''
Phys.\ Lett.\ B {\bf 490}, 87 (2000)
[arXiv:hep-ph/0007220].
%%CITATION = PHLTA,B490,87;%%

\bibitem{top-bottom}
%\cite{Datta:1997us}
%\bibitem{Datta:1997us}
A.~Datta, J.~M.~Yang, B.~L.~Young and X.~Zhang,
%``Effects of R-parity-violating supersymmetry in single top production at
%the Tevatron,''
Phys.\ Rev.\ D {\bf 56}, 3107 (1997)
[arXiv:hep-ph/9704257];
%%CITATION = PHRVA,D56,3107;%%
%\cite{Oakes:1997zg}
%\bibitem{Oakes:1997zg}
R.~J.~Oakes, K.~Whisnant, J.~M.~Yang, B.~L.~Young and X.~Zhang,
%``Single top quark production as a probe of R-parity-violating SUSY at p p
%and p anti-p colliders,''
Phys.\ Rev.\ D {\bf 57}, 534 (1998)
[arXiv:hep-ph/9707477];
%%CITATION = PHRVA,D57,534;%%
%\cite{Chiappetta:1999cd}
%\bibitem{Chiappetta:1999cd}
P.~Chiappetta, A.~Deandrea, E.~Nagy, S.~Negroni, G.~Polesello and J.~M.~Virey,
%``Single top production at the LHC as a probe of R parity violation,''
Phys.\ Rev.\ D {\bf 61}, 115008 (2000)
[arXiv:hep-ph/9910483].
%%CITATION = PHRVA,D61,115008;%%

%\cite{Hahn:1998yk}
\bibitem{Hahn:1998yk}
T.~Hahn and M.~Perez-Victoria,
%``Automatized one-loop calculations in four and D dimensions,''
Comput.\ Phys.\ Commun.\ {\bf 118} (1999) 153
[arXiv:hep-ph/9807565].
See also http://www.feynarts.de/looptools/
%%CITATION = HEP-PH 9807565;%%

%\cite{vanOldenborgh:1989wn}
\bibitem{vanOldenborgh:1989wn}
G.~J.~van Oldenborgh and J.~A.~Vermaseren,
%``New Algorithms For One Loop Integrals,''
Z.\ Phys.\ C {\bf 46} (1990) 425.
%%CITATION = ZEPYA,C46,425;%%

\end{thebibliography}
\end{document}